\documentstyle[12pt]{article}
\begin{document}
\begin{center}

{\large\bf Common physical mechanism  for integer and fractional
quantum Hall effects} \vskip 1cm Jianhua
Wang$^{1}$~~Kang Li$^{2}$~~Shuming Long $^{1}$~~Yi Yuan $^{1}$
\\\vskip 1cm

{\it\small 1.~~ Shaanxi University  of Technology,
Hanzhong, 723001, P.R. China

2.~~ Hangzhou Normal University,
Hangzhou, 310036, P.R. China \\}
\end{center}

\begin{abstract}

Integer and fractional quantum Hall effects were studied with different physics models and explained by different physical mechanisms.  In  this paper, the common physical mechanism for integer and fractional quantum Hall effects is studied, where  a new unified formulation of integer and fractional quantum Hall effect is presented. Firstly, we introduce a 2-dimensional ideal electron gas model in the presence of  strong magnetic field with symmetry gauge, and the transverse electric filed $\varepsilon_2$ is also introduced to balance Lorentz force. Secondly, the  Pauli equation is solved where the wave function and energy levels is given explicitly. Thirdly, after the calculation of the degeneracy density for 2-dimensional ideal electron gas system, the Hall resistance of the system is obtained, where the quantum Hall number $\nu$ is introduced. It is found that the new defined $\nu$ , called filling factor in the literature, is related to radial quantum number n and angular quantum number $|m|$, the different $n$ and $|m|$ correspond to different $\nu$. This provides unification explaination for integer and fractional quantum Hall effects. It is predicated that more new cases exist of fractional quantum Hall effects without the concept of fractional charge.

\noindent PACS number(s): 2.20.My, 71.45.-d, 73.40.Lq

\noindent Keywords:  integer quantum Hall effect, fractional quantum Hall effect, Pauli equation, quantum Hall number
\end{abstract}

\section{Introduction}

In 1879, E. H. Hall discovered that when a conductor carrying an
electric current perpendicular to an applied magnetic field develops
a voltage gradient which is  transverse to both the current and the
magnetic field. This phenomenon is called Hall effect. About 100 years later,  Klaus von Klitzing, in
1980, made the unexpected discovery that, under low temperature and strong magnetic field, the Hall conductivity was
exactly quantized\cite{klizing1}, in which the Hall conductivity $\sigma$ takes on the
quantized values, i.e. $\sigma =\nu\frac{e^2}{h}$ and $\nu$ takes integer values,  we call it quantum Hall effect (QHF).
For this finding, von Klitzing was awarded the 1985 Nobel Prize in Physics\cite{klizing2}.
Very soon after that, under much more low temperature,
the fractional quantum Hall effect(FQHE) was experimentally
discovered in 1982 by Daniel Tsui and Horst Stormer\cite{tsui1}, in which
$\nu$, called filling factor, takes fractional values. Each particular value of
the magnetic field corresponds to %(the ratio of electrons to magnetic flux quanta)%
$\nu=\frac{p}{q}$, where p and q are integers with no common
factors. Here q turns out to be an odd number with the exception of
two $\nu $'s $ 5/2$ and $7/2$. The principal series of such
fractions are
$\frac{1}{3}, \frac{2}{5}, \frac{3}{7},$ etc.,and $\frac{2}{3}, \frac{3}{5}, \frac{4}{7},$ etc..
All fractions have an odd denominator.  The effect was explained by
Laughlin in 1983, using a novel quantum liquid phase that accounts
for the effects of interactions between electrons. Tsui, Stormer,
and Laughlin were awarded the 1998 Nobel Prize in physics for their work.

The theoretical study of integer and fractional quantum Hall effects has lasted about thirty years and has great influence on physics\cite{klizing3}. For examples,
the theory of Laughlin which introduced several novel
concepts in correlated quantum fluids, inspired analogous
effects in other subfields of physics; the quantum Hall effect was generalized
to four dimensions \cite{zhang} in order to study the
¡°interplay between quantum correlations and dimensionality
in strongly correlated systems¡±; two-dimensional
electron systems were modeled by strings interacting with
D-branes \cite{BSTB}, where the fractionally-charged quasi-particles
and composite fermions were described in the language
of string theory;  an interesting analogy between
the quantum Hall effect and black hole has been reported, and in particular,
the edge properties of a quantum Hall effect system have been
used to model black hole physics from the point of view
of an external observer \cite{BMC}. Important developments of the quantum Hall effect have also taken place from the theoretical point of view\cite{ezawa}-\cite{VJP}.

However, in the most studies, integer and fractional quantum Hall effects were studied with different physics models and explained by different physical mechanisms. In this paper, an alternative unification description for integer and fractional quantum Hall effects is given, in which pure quantum mechanics theory is used and concepts of the fractionally-charged quasi-particles and composite fermions are not necessary.

\section{ The Hall density of electron gas model under strong magnetic field}

In this section, we study the wave function and energy levels for electrons moving in an electromagnetic field, and give the expectation value of the electron moving size. After a new definition of quantum Hall number $\nu$ , we give the Hall density explicitly.

The Dirac equation for an electron moving in electromagnetic filed is usually expressed as,

\begin{equation}\label{Dirac}
    i\hbar\frac{\partial}{\partial t}\psi
    =[c\vec{\alpha}\cdot(\vec p+\frac{e}{c}\vec A)-e\phi+\mu c^2\beta]\psi .
\end{equation}

Let's consider the situation of the electron moving in $x-y$ plane, and uniform magnetic field applied in $z$-direction. With the symmetric gauge, i.e. $\vec A = (-\frac{B}{2}y, \frac{B}{2}x, 0)$, and the scale potential of electric field is $\phi =-\varepsilon_1 x-\varepsilon_2 y $, in which $\varepsilon_1$ is the longitudinal external field and $\varepsilon_2$ is the transverse electric filed which can balance Lorentz force, then the Pauli equation for a single electron moving in an external electromagnetic field is simplified as

\begin{equation}\label{Pauli}
    i\hbar\frac{\partial}{\partial t}\psi =(H^0+H')\psi
\end{equation}
where
\begin{equation}\label{H}
  \begin{array}{ll}
  H^0&=\frac{1}{2\mu}(\hat{p}_x -\frac{eB}{2c}y)^2+\frac{1}{2\mu}(\hat{p}_y +\frac{eB}{2c}x)^2+\frac{1}{2\mu}(\hat{p}_z)^2+\frac{e\hbar B}{2\mu c}\sigma_z,\\
  H'&=e\varepsilon_1 x+e\varepsilon_2 y
  \end{array}
\end{equation}
the last term of $H^0$ is the Stern-Gerlach term.

Treating $H'$ as a perturbative term, then in cylinder coordinate description, we find, after tedious calculation, the perturbative energy levels and the stationary wave function are as follows.

\begin{equation}\label{energylevel}
 E_{N\lambda}
    =\frac{1}{2\mu}p_{z}^2+(N+\frac{\lambda+1}{2})\frac{e\hbar B}{2\mu c}-\frac{\mu c^2(\varepsilon_1^2+\varepsilon_2^2)}{2B^2},
 \end{equation}

\begin{equation}\label{wavefunction}
    \psi_{N m\lambda}(\rho,\varphi,z,s,t)=\phi_{N,m}(\xi,\varphi)\chi_\lambda(s)e^{-ip_z z/\hbar}e^{-iEt/\hbar},
\end{equation}
where
\begin{equation}
\begin{array}{lr}
 \phi_{N,m}(\xi,\varphi)=\phi^{(0)}_{N,m}-\frac{\alpha}{4}(\eta-i)\sqrt{N+1}\phi^{(0)}_{N+1,m+1}(\xi,\varphi)&~\\
 + \frac{\alpha}{4}(\eta+i)\sqrt{N}\phi^{(0)}_{N-1,m-1}(\xi,\varphi),&~\\
 N=n+\frac{|m|+m}{2}, ~ n=0,1,2,\cdots ; m=0\pm 1\pm 2 \cdots ; \lambda =-1, 1,&~
 \end{array}
\end{equation}
 and
 \begin{equation}
 \phi^{(0)}_{N,m}=
     [\frac{(n+|m|)!}{a^2\pi n!}]^{1/2}e^{im\varphi}e^{-\xi^{2}/2}
   \Sigma_{k=0}^{n}
    \frac{(-1)^k[\begin{array}{c}n\\k\end{array}]}{(k+|m|)!}
    \xi^{2k+|m|},
     \end{equation}
 in which $n$ is radial quantum number, $m$ is angular momentum quantum number along $z$ direction, $\lambda$ is the spin quantum number,  $a=(2\hbar c/eB)^{1/2}$, $\xi=\frac{\rho}{a},~ \eta =\varepsilon_1/\varepsilon_2, \alpha =\frac{\mu e\varepsilon_2}{\hbar^2}(\frac{2\hbar c}{eB})^{3/2}$ and
$[\begin{array}{c}n\\k\end{array}]$ is the binomial coefficient.

Now, let's calculate the expectation value of the electron moving size $\pi <\rho^2>$. Obviously, the expectation value does not depend on the factors $\chi_\lambda(s)e^{-ip_z z/\hbar}e^{-iEt/\hbar}$ of the wave function, but depends on the function of $ \phi_{N,m}(\xi,\varphi)$. Namely,

\begin{equation}\label{expectation}
\begin{array}{ll}
    <S>_{N m}=<\pi\rho^2>_{N m}=\pi a^2\int_0^{2\pi}d\varphi\int_0^\infty |\phi (\xi, \varphi)|^2\xi^3d\xi&~\\
    =(2N+1-m)[1+(2N+1)\beta]\frac{hc}{eB}+\beta\frac{hc}{eB},&~
\end{array}
\end{equation}
where $\beta =\mu c^2(\varepsilon_1^2+\varepsilon_2^2)/2B^2$.This shows that the expectation value of electron moving size depends on the state quantum number  $N$,$m$ and the external magnetic filed.

The quantum Hall effect is the quantum effect when the magnetic filed is very strong and the temperature is very low. In this condition the electrons will be fully polarized, and quantum number $\lambda =-1$, and the $m\leq 0$.  Because $\beta\ll 1$, so when  $n$ and $m$ are not very big, equation (\ref{expectation}) can be written as

\begin{equation}\label{expectation2}
\begin{array}{ll}
<S>_{N m}=(2n+|m|+1)[1+(2n +m +|m|+1)\beta]\frac{hc}{eB}+\beta\frac{hc}{eB}&~\\
\approx (2n+|m|+1)\frac{hc}{eB}.&~
\end{array}
\end{equation}

When the energy level is chosen, the number of electron states is determined by quantum number $m$. In the region of $<S>_{N m}$, angular quantum number can be $0, -1, -2, \cdots , -|m|$ , i.e. $|m|+1$ values,   so the energy degeneracy density (energy degeneracy on per unit area) for electron gas in strong magnetic field is
\begin{equation}\label{nB}
    n_B=\frac{|m|+1}{<S>_{N m}}=\frac{|m|+1}{2n +|m|+1}\frac{eB}{hc},
\end{equation}
in which $n_B$ is also called Hall density.

Now let's define a quantum Hall number as
\begin{equation}\label{nu}
    \nu=\frac{|m|+1}{2n +|m|+1},
\end{equation}
 called filling factor in the literature, then the Hall density has the form of
\begin{equation}\label{nB2}
    n_B=\nu\frac{eB}{hc}.
\end{equation}
This shows that Hall density is proportional to the strength of magnetic field.

\section{Physical understanding of ideal electron gas model under the strong magnetic field}

The conditions for quantum Hall effect are low temperature and strong magnetic field. When an electron moves in a magnetic field with the field strength $B=200000$ Gauss, the interaction energy between electron and magnetic field is $\varepsilon_0 =e\hbar B/\mu c= 0.00232$eV. Though the energy is very small, it is much bigger than the dynamic energy of electron with velocity $v=3\times 10^{5}$cm/s, $p_z^2/(2\mu)=2.56\times 10^{-5}$eV. So, in the presence of strong magnetic field, the electron moving size,  in ground state, is $150$ times than that of the electron in the ground state of hydrogen atom. Therefore, in strong magnetic field, the electrons have much freedom than they are in atoms, but the interaction between magnetic field and electron is much smaller than the interaction of electron with nuclear of atom. Thus, when the Lorenze force is balanced by a transverse electric filed $\varepsilon_2$ , the dilute electron gas in magnetic field can be considered as ideal electron gas system.

On the other hand, the interaction among electrons is materialized from the electric field $\varepsilon_2$ in $y$ direction, which is produced by the nonuniform distribution of electrons along $y$ direction. According to (\ref{expectation2}), the expectation value of the electron moving size proportional to $1/B$, that is, bigger $B$ to corresponding less expectation value of the electron moving size. For the fractional quantum Hall effect, $B$ is much stronger, and then expectation value of the electron moving size get extreme less, and less interaction among electrons. So in the situation of integer quantum Hall effect and fractional quantum Hall effect, the ideal electron gas model is reasonable.

After some calculations, we get the expectation value of electric currents as

\begin{equation}\label{currents}
\begin{array}{ll}
 J_x&=<j_x>=-en_s<v_x>=en_s\frac{c\varepsilon_2}{B},\\
 J_y&=<j_y>=-en_s<v_y>=-en_s\frac{c\varepsilon_1}{B}.
\end{array}
\end{equation}

From the equations above, we can understand why there exists superconductivity in $x$ direction on Hall plateau.

Since we can define
\begin{equation}
    \frac{n_s}{n_B}=i,
 \end{equation}
and

\begin{equation}
    w=i\nu,
 \end{equation}
from the definition of Hall resistance and equation (\ref{nB}), we get the Hall resistance as
 \begin{equation}\label{ruo}
    \rho_{xy}=-\rho_{yx}=\frac{B}{n_s ec}=\frac{1}{i\nu}\frac{h}{e^2}=\frac{1}{w}\frac{h}{e^2}.
 \end{equation}

 Obviously, the Hall resistance only depends on the quantum Hall number ¦Í, and the later depends on the ratio of $|m|+1$ and $2n +|m|+1$ . This is the key result of this paper. With this result, the quantum Hall effect and fractional quantum Hall effect can be unified formulated.

 \section{Unification formulation of integer and fractional quantum Hall effects}

In this section, we will use the results above to provide a unification description for both integer and fractional quantum Hall effects. From the formula of quantum Hall number (11), we can see that different quantum number n and m correspond to different quantum states, and that $\nu$ can take value 1 and also fraction, they relate respectively to integer quantum Hall effect and fractional quantum Hall effect.

When $n=0$ , from equation (11) we get $\nu=1$ . And now the angular quantum number can still take one of  $|m|+1$ values: $0, -1,-2 \cdot\cdot\cdot-|m|$. In other words, on one energy level, the electrons of the electron gas system can fill different state with different angular quantum number $0, -1,-2 \cdot\cdot\cdot-|m|$. The less absolute value of angular quantum number correspond to the more stable state and the less moving range. So, all electrons firstly occupy the state of angular quantum number equivalent to 0, and when the groud state is full filled, Hall plateau appears. When the magnetic field becomes less, electron density of electron gas also gets less, then the extra electrons of state $m=0$ will fill the state of $m=-1$. When these states are full filled, the Hall plateau appears again. When the extern magnetic field gets less and less, the electrons will fill the states of $m=-2,-3 \cdot\cdot\cdot$, other Hall plateaus appear one by one. This is the integer Hall effect.

When  $n \neq 0$, the electrons of the electron gas system stay in excited sates. Different n and different $|m|$ correspond to different fractional quantum Hall number and thus correspond to different Hall effects, which is just the fractional Hall effect. Therefore, the fractional quantum Hall effect corresponds to the filling of electrons to exciting states. The principal of the filling is the same as that of the integer quantum Hall effect filling the electrons to ground sates. Hall plateaus also appear when the corresponding state of the chosen angular quantum number is full filled by electrons.
Now, Table I shows possible value of Hall quantum numbers $\nu=\frac{|m|+1}{2n +|m|+1}$,  where $n =0,1,2 \cdot\cdot\cdot,9$ and $m = 0,-1,-2 \cdot\cdot\cdot,-9$.

\begin{table}
  \centering
  \caption{quantum Hall numbers when $n =0,1,\cdot\cdot\cdot,9$ and $m = 0,-1,\cdot\cdot\cdot,-9$ }\label{table11}
  \begin{tabular}{|c|c|c|c|c|c|c|c|c|c||c|}
    \hline
   $ m\backslash ~n$ &0 & 1 &2 & 3 & 4 & 5 & 6 & 7 & 8 & 9  \\
    \hline
   0 &1 &1/3   &1/5    &1/7    &1/9    &1/11   &1/13   &1/15   &1/17   &1/19   \\
    \hline
   -1 &1 &1/2  &1/3    &1/4    &1/5    &1/6    &1/7    &1/8    &1/9    &1/10     \\
    \hline
   -2 &1  &3/5  &3/7    &1/3    &3/11   &3/13   &1/5    &3/17   &3/19   &1/7      \\
    \hline
   -3 & 1 &2/3  &1/2    &2/5    &1/3    &2/7    &1/4    &2/9    &1/5    &2/11    \\
    \hline
   -4 &1  &5/7  &5/9    &5/11   &5/13   &1/3    &5/17   &5/19   &5/21   &5/23     \\
    \hline
   -5 &1  &3/4  &3/5    &1/2    &3/7    &3/8    &1/3    &3/10   &3/11   &1/4      \\
    \hline
   -6 &1  &7/9  &7/11   &7/13   &7/15   &7/17   &7/19   &1/3    &7/23   &7/25    \\
    \hline
   -7 &1 &4/5  &2/3    &4/7    &1/2    &4/9    &2/5    &4/11   &1/3    &4/13     \\
    \hline
   -8 &1 &9/11 &9/13   &3/5    &9/17   &9/19   &3/7    &9/23   &9/25   &1/3    \\
    \hline
  -9 &1 &5/6   &5/7    &5/8    &5/9    &1/2    &5/11   &5/12   &5/13   &5/14     \\
     \hline
 \end{tabular}
\end{table}

According to Table I and the analysis above, some remarks are made as follows.

1). The  $\nu$  of all known fractional quantum Hall effects are concluded in the table. For examples, $\nu= \frac{1}{3},\frac{2}{3}, \frac{2}{5},  \frac{3}{5},\frac{1}{7},\cdot\cdot\cdot$, of which $\nu= \frac{1}{3}$  appears firstly in the table and also most frequently. From this point of view, its first discovering by experiment is easy understood, and also it corresponds to the excited state with the lowest energy.

2).  $\nu= \frac{1}{2}$  also appears frequently in Table I. Its first corresponding quantum number is $n=1,m=-1$ ; the second is $n=2,m=-3$; its other corresponding quantum number is $n=|m|-1$. While, no Hall plateaus appear corresponding to $\nu= \frac{1}{2}$ , which needs further studying.

3). When $m=0$, the Hall quantum number is reduced to $\nu=1/(2k+1)$ , and the electron gas system in these states are so called incompressible quantum liquid. In this case, electrons are in s-wave states, namely, in circular Bohr orbit.

4). Besides the quantum Hall numbers defined in (11) which fully described the known fractional Hall effect, the fractional Hall effect for other quantum Hall numbers may also exist. We expect they can be experimentally checked very soon.

5). From equation (15) we know when $n_s/n_B=i>1$  , the quantum Hall numbers can be obtained for improper fraction. For example, $n_s/n_B=i=5, \nu=1/2$ is the fractional quantum Hall effect for $\nu=5/2$ .

\section{Conclusion remarks }

The electron gas model is used in this paper to unify the integer and fractional quantum Hall effect, in brief, with the solution of the Schrodinger equation, we firstly got the wave function, moving size, and the expectation value of current density for the single-electron in strong magnetic filed and weak electric filed.  Explicit calculations show that  $B$, external magnetic field, is in proportion to Hall density $n_B$,  but in inverse proportion to the expectation value of electron moving size $<S>_{N m}$,   all electrons have almost the same possibility to meet with each other in both the conditions where the integer quantum Hall effect and the fractional quantum effect appear. This is the physical foundation of this paper where we use single electron model to describe the fractional quantum Hall effects.  With $\nu=\frac{|m|+1}{2n +|m|+1}$, a unified interpretation is given to the integer and fractional quantum effects. This paper shows both effects have common physical mechanisms as follows. (1)Every electrons in dilute two-dimensional electron gas system is same as one single electron. (2) Energy levels for electrons with angular quantum number $m\leq 0$ is not related to  $m$, so the  energy level degeneracy is infinite theoretically, but in limited range, the degeneracy is finite.  (3) Concepts of fractionally-charged quasi-particles and composite fermions are not used in this study.  The electron gas model need to be further studied, for example, the  energy levels and wave function for the electron gas system etc. will be reported in our forthcoming  papers.

\section{Acknowledgments} This work is supported by the National
Natural Science Foundation of China (11147181, 10875035 and 10665001),
an open topic of the State Key Laboratory for Superlattices and Microstructures (CHJG200902), the scientific research project in Shaanxi Province (2009K01-54) and Natural Science Foundation of Zhejiang Province (Y6110470).
 The authors are also grateful to the support from the KITPC, Beijing, China.


\begin{thebibliography}{10}

\bibitem{klizing1}
K. von Klitzing, G. Dorda, M. Pepper, New Method for
High-Accuracy Determination of the Fine-Structure Constant Based on
Quantized Hall. Physical Review Letters  45, 494¨C497 (1980)
\bibitem{klizing2}
  Klaus von Klitzing, Nobel Lecture: The quantized Hall e?ect.  Rev. Mod. Phys. 58, 519-531 (1986).
\bibitem{tsui1}
D.C. Tsui, H.L. Stormer, A.C. Gossard, Two-Dimensional
Magnetotransport in the Extreme Quantum Limit. Physical Review
Letters 48, 155(1982)
\bibitem{laughlin1}
R.B. Laughlin, Anomalous Quantum Hall Effect: An
Incompressible Quantum Fluid with Fractionally Charged Excitations.
Physical Review Letters 50, 1395(1983)
\bibitem{klizing3}
Tapash Chakraborty, Klaus von Klitzing, Taking stock of the quantum Hall effects: Thirty years on.arxiv:1102.5250v1.
\bibitem{zhang}
S.-C. Zhang and J. Hu, A four-dimensional generalization
of the quantum Hall effect. Science 294, 823-828(2001).
\bibitem{BSTB}
B.A. Bernevig, L. Susskind, N. Toumbas, and J.H.
Brodie, How Bob Laughlin tamed the giant graviton
from Taub-NUT space. JHEP02, 003 (2001).
\bibitem{BMC}
A.P. Balachandran, A. Momen, and L. Chandar, ¡°Edge
states in gravity and black hole physics¡±, Nucl. Phys.
B 461, 581-596 (1996); Y.S. Myung, BTZ black hole
and quantum Hall effect in the bulk-boundary dynamics.
Phys. Rev. D 59, 044028 (1999).
\bibitem{ezawa}
Z.F. Ezawa, Quantum Hall Effects: Field Theoretical Approach
and Related Topics (World Scientific, Singapore,
2008), 2nd edition.
%\bibitem{lwl}
%Shuming Long, kang Li and Jianhua Wang, The theoretical calculation for new explaination of integer and quantum Hall effect, to be submitted for publication (2012).
\bibitem{DGSGHUM}
M. Dolev, Y. Gross, R. Sabo, I. Gurman, M. Heiblum, V. Umansky, and D. Mahalu ,¡°Characterizing Neutral Modes of Fractional States in the Second Landau Level¡± £¬Phys. Rev. Lett. 107, 036805 (2011)£»
\bibitem{GZCBK}
F. Ghahari, Y. Zhao, P. Cadden-Zimansky, K. Bolotin and P. Kim, ¡°Measurement of the ¦Í = 1/3 fractional quantum Hall energy gap in suspended graphene¡±, Phys. Rev. Lett. 106, 046801 (2011)£»
\bibitem{VJP}
V. Venkatachalam, A. Jacoby, L. Pfei?er, and West, ¡°Local charge of the ¦Í = 5/2 fractional quantum Hall state¡±, Nature 469, 185-188 (2011).


\end{thebibliography}
\end{document}